\documentclass[aps,twocolumn,amsfonts,amssymb,amsmath,pra,showpacs]{revtex4}

\usepackage{graphicx}
\usepackage{epsfig}

\newcommand{\field}[1]{\mathbb{#1}}

\newcommand{\comment}[1]{}

\begin{document}

\author{Karel Lemr}
\author{Jarom\'{\i}r Fiur\'a\v{s}ek}
\affiliation{Department of Optics, Palack\'y University, 17. listopadu 50, 77200 Olomouc, Czech Republic}
\title{Preparation of entangled states of two photons in several spatial modes}

\begin{abstract}
We describe a protocol capable of preparing an arbitrary state of two photons in several spatial modes using pairs of photons generated by spontaneous parametric down-conversion, linear optical elements and single-photon detectors or post-selection. The protocol involves unitary and non-unitary transformations realizable by beam splitters and phase shifters. Non-unitary transformations are implemented by attenuation filters. The protocol contains several optimization capabilities with the goal of improving overall probability of its success. We also show how entangled two-photon states required for quantum computing with linear optics can be prepared using a very simple and feasible scheme.
\end{abstract}

\pacs{03.67.Bg, 42.50.Dv}

\maketitle

\section{Introduction}
Quantum information processing (QIP) has demonstrated an important development in recent years \cite{bib:alber:quantum_inform}. The significant advances in implementing various protocols for QIP can, in the future, lead to long-distance quantum communication \cite{bib:bowmeester:quantum_inform} or creation of a quantum computer \cite{bib:nielsen:quantum_comput}. Among many experimentally tested physical platforms, optical implementations appear particularly promising. Light is an ideal carrier of quantum information, as it can be transmitted over long distances through optical fibers or even in a free space. A crucial prerequisite for optical quantum information processing is the ability to generate and manipulate various highly non-classical and entangled states of light beams. Entangled states of light are necessary for instance for quantum teleportation \cite{bib:benett:teleport,bib:bouwmeester:teleport,bib:braunstein:teleport,bib:furusawa:unconditional}, quantum gates in quantum computing with linear optics \cite{bib:pittman:circuit,bib:kiesel:cphase,bib:gasparoni:cnot,bib:brien:cnot,bib:pittman:logic}, or quantum repeaters \cite{bib:briegel:repeaters}.

These wide potential applications motivate the effort aimed at developing and demonstrating preparation protocols for various interesting quantum states of light. Single and two-photon Fock states were conditionally prepared from two-mode squeezed vacuum by measuring the number of photons in the idler mode \cite{bib:lvovsky:reconstruction,bib:zavatta:tomographic_photon,bib:ourjoumtsev:tomography_two_photons,bib:nielsen:photon_number_states}. Schr\"{o}dinger cat-like states were generated by subtracting a single photon from pulsed or continuous squeezed beams \cite{bib:dakna:kitten,bib:ourjoumtsev:kitten,bib:neergaard-nielsen:kitten,bib:wakui:photon_substracted,bib:ourjoumtsev:kitten_from_fock}. Photon-added coherent states were produced experimentally using down-conversion seeded with a weak coherent signal and conditioning on detection of a photon in the idler mode \cite{bib:zavatta:single_add}. 
Schemes for preparation of arbitrary single mode states via repeated addition or subtraction of single photons have been suggested \cite{bib:dakna,bib:fiurasek:singlemode_fockstates}.  A universal procedure for conditional preparation of arbitrary multimode states of light with linear optics, single photons, coherent states and single-photon detectors has been developed \cite{bib:fiurasek:multimode_fockstates}.

A lot of attention has been paid to creation of two-mode N-photon entangled states, the so-called NOON states, \cite{bib:mitchell:super,bib:kok:creation,bib:fiurasek:nphoton,bib:lee:linear,bib:hofmann:generation,bib:nielsen:noon,bib:cable:noon,bib:kapale:noon}. It was shown theoretically and demonstrated experimentally that such states can be conditionally generated from input single-photon states by multiphoton interference in a properly tailored interferometer. In a similar way, projection onto arbitrary NOON-type state can be accomplished \cite{bib:sun:noon}. 
The N-photon entangled states can find applications not only in quantum information processing, but can be also used  for ultra-precise measurements \cite{bib:yurke:input,bib:hillery:interfer,bib:brif:nonclassical,bib:dowling:inputport,bib:resch:ultra-precise_measurements} or quantum lithography \cite{bib:boto:litography,bib:bjork:litography,bib:dangelo:twophoton}. For advanced applications, multimode entangled states will be required. However, the technique used for preparation of arbitrary two-mode N-photon states \cite{bib:mitchell:super} cannot be immediately extended to multimode case. 

\begin{figure}[!t!]
\begin{center}
\scalebox{0.9}{\includegraphics{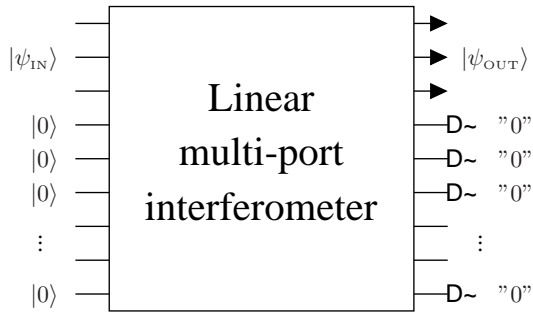}}
\caption{General multiphoton interferometer transforming a three-mode input state $|\psi_{\mathrm{in}} \rangle$ into $|\psi_{\mathrm{out}} \rangle$ using only vacuum ancillae. We post-select cases when both photons are present in the first three output modes.}
\label{fig:general_interferometer}
\end{center}
\end{figure}

In this paper, we show that we can exploit experimentally accessible two-photon entangled states generated in the process of spontaneous parametric down-conversion to extend the group of states that can be prepared using only passive linear optics and single-photon detectors. We propose a feasible protocol for generation of an arbitrary quantum state of two photons in three optical spatial modes
\begin{eqnarray}
|\psi_{\mbox{\footnotesize{target}}}\rangle &=& \alpha |200\rangle + \beta |020\rangle + \gamma |002\rangle 
\nonumber \\
& &  + \delta |110\rangle + \epsilon |101\rangle + \eta |011\rangle,
\label{eq:target}
\end{eqnarray}
where the numbers in each bracket denote the number of photons in the first, second and third mode, respectively. To prepare an arbitrary state (\ref{eq:target}) we use a multiphoton interferometer, which is schematically shown in Fig. \ref{fig:general_interferometer}. A two-mode 
two-photon state can be prepared in this way from a separable state $|110\rangle$ as the input. This approach however fails if the number of modes becomes higher. We will show that using an entangled input state permits us to create arbitrary states of the form (\ref{eq:target}). We will then generalize our procedure to conditional generation of arbitrary multi-mode entangled two-photon state. The scheme is cheap in terms of required resources because it involves only two photons and vacuum ancillae which is a great practical advantage.

The rest of the present paper is organized as follows. In Sec. II we prove that separable input two-photon state is insufficient for generation of arbitrary multimode two-photon state via interference and post-selection. In Sec. III we describe the scheme for generation of arbitrary three-mode states (\ref{eq:target}). Results of numerical simulations of generation of various states are presented in Sec. IV. In Sec. V we generalize the state preparation procedure to arbitrary number of modes.
A simple experimentally feasible scheme tailored for generation of a specific class of four-mode two-photon states, the so-called KLM states, is discussed in Sec. VI. Finally, Sec. VII contains a brief summary of the main results and conclusions.

\section{Motivation}

First of all we prove that a separable input state is not sufficient for preparation of an arbitrary state of the form (\ref{eq:target}) by the interferometric method sketched in Fig.~\ref{fig:general_interferometer}. We express the input state $|110\rangle$ as $\hat{a}_1^\dag \hat{a}_2^\dag |\mathrm{vac}\rangle$, where $\hat{a}_i^\dag$ denotes creation operator for the $i^{\mbox{\footnotesize th}}$ mode and $|\mathrm{vac}\rangle$ stands for the vacuum state. General interferometer performs linear transformation that can be written as
\begin{equation}
\hat{a}_{i, \mathrm{in}}^\dag = \sum_{j}u_{ij} \hat{a}_{j, \mathrm{out}}^\dag,
\label{eq:a_general_transform}
\end{equation}
where $u_{ij} \in \field{C}$ are elements of a unitary matrix $U$. For future reference, we note that  the interference of modes $a$ and $b$ on  a beam splitter BS with amplitude transmittance $t$ and reflectance $r=\sqrt{1-t^2}$ is governed by
\begin{eqnarray}
a^{\dagger}_{\mathrm{out}} = t a^{\dagger}_{\mathrm{in}}+rb^{\dagger}_{\mathrm{in}},
\qquad
b^{\dagger}_{\mathrm{out}} = t b^{\dagger}_{\mathrm{in}}-ra^{\dagger}_{\mathrm{in}},
\end{eqnarray}
and a phase shifter PS shifts the phase of a single mode according to $a^\dagger_{\mathrm{out}}=e^{i\phi} a^\dagger_{\mathrm{in}}$. The separable state $|110\rangle$ will be transformed as follows,
\begin{equation}
|110\rangle \rightarrow \left (\sum_{j} u_{1j} a_{j, \mathrm{out}}^\dag \right ) \left ( \sum_{k} u_{2k} a_{k, \mathrm{out}}^\dag \right ) |\mathrm{vac}\rangle.
\label{eq:factorised}
\end{equation}
Conditional projection of all output ancillae onto vacuum then gives
\begin{equation}
|110\rangle \rightarrow \left (\sum_{j=1}^3 u_{1j} a_{j, \mathrm{out}}^\dag \right ) 
\left ( \sum_{k=1}^3 u_{2k} a_{k, \mathrm{out}}^\dag \right ) |\mathrm{vac}\rangle.
\label{eq:factorised_cond}
\end{equation}
It is easy to verify that the class of achievable target states (\ref{eq:factorised_cond}) is limited. For example it can be shown that the state
\begin{equation}
|110\rangle + |101\rangle + |011\rangle = (a^\dag b^\dag + a^\dag c^\dag + b^\dag c^\dag)|\mbox{vac}\rangle
\label{eq:unobteinable}
\end{equation}
cannot be expressed in the factorized form  (\ref{eq:factorised_cond}) and is therefore not obtainable from separable state like $|110\rangle$ by multiphoton interference using only vacuum ancillae. Addition of single photon ancillae would make the preparation possible, but also experimentally more difficult \cite{bib:vanmeter:general}.

\begin{figure}[!t!]
\begin{center}
\scalebox{1}{\includegraphics{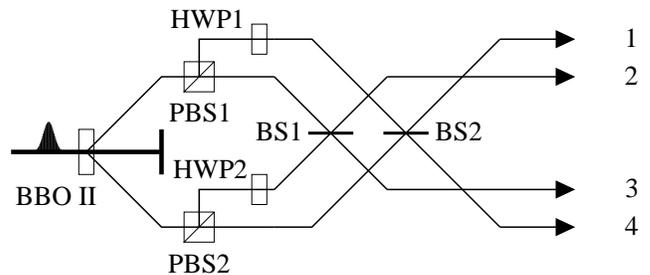}}
\end{center}
\caption{Initialization phase of the protocol. Pair of photons obtained by spontaneous parametric down-conversion in a nonlinear crystal BBO is sent to polarizing beam splitters PBS1 and PBS2 which transmit horizontally polarized photons and reflect  vertically polarized photons.
Half wave plates HWP1 and HWP2 are inserted to modes 2 and 3 to align the polarization of all beams to vertical. The photons then interfere on two ordinary balanced beam splitters BS1 and BS2.}
\label{fig:init}
\end{figure}

\begin{figure*}[!t!]
\begin{center}
\includegraphics{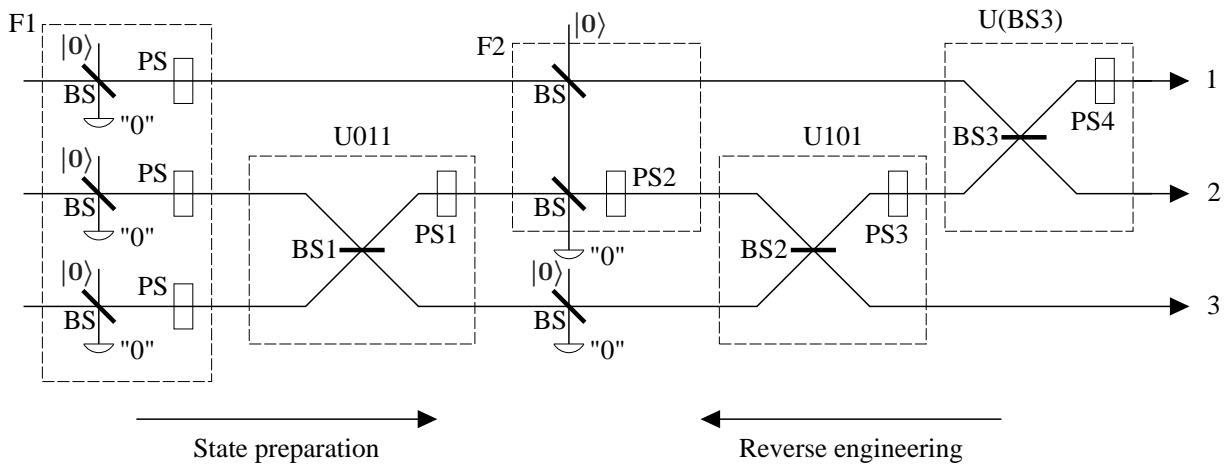}
\end{center}
\caption{Preparation phase. The  linear interferometer involves three unitary operations 
U011, U101 and U(BS3) composed of beam splitters BSj and phase shifters PSk. The preparation also requires two attenuation filters F1 and F2 which are implemented using unbalanced beam splitters and vacuum ancillae. Successful filtration can be verified by monitoring the output ancilla modes with single-photon detectors. In the design of the scheme we imagine 
reverse propagation of the state backwards from the output to the input.}
\label{fig:scheme}
\end{figure*}

Even with one fixed entangled state as the input, one can not prepare arbitrary state (\ref{eq:target}) using only deterministic unitary transformations. To prove this, let us consider two photons in $n$ optical modes. Such quantum state is then determined by
\begin{equation}
2 R_2^{n} - 2 = n^2 + n - 2
\label{eq:state_param}
\end{equation}
real parameters where $R_k^n={n+k-1 \choose k}$ denotes the number of combinations with repetitions of $n$ elements in $k$ classes. The term $-2$ corrects the number of parameters due to normalization and irrelevant overall phase of the state. On the other hand, $n$-dimensional special unitary transformation represented by $n \times n$ complex matrix contains
\begin{equation}
2 n^2 - n -2 C_2^n -1  = n^2 - 1
\label{eq:unit_param}
\end{equation}
real parameters, where $C_k^n= {n \choose k}$ denotes the number of combinations without repetitions of $n$ elements in $k$ classes.  The factor $-n$ in Eq. (\ref{eq:unit_param}) accounts for the reduction of the number of free parameters due to normalization condition of each row of the matrix. The term $-2 C_2^n$ represents correction due to condition of orthogonality of rows and the factor $-1$ occurs since  the determinant of a matrix belonging to the  $\mathrm{SU}(n)$ group  is fixed and equal to $1$. It is evident that for any $n > 1$ the number of  parameters specifying  the quantum state is larger than the number of degrees of freedom allowed by unitary transformation. This results in the need of probabilistic non-unitary filtering operations 
in a fully general preparation scheme.

\section{Preparation scheme}

Here we show that by exploiting the two-photon entangled states generated by spontaneous parametric down-conversion as a resource, we can conditionally generate arbitrary state of the form  (\ref{eq:target}) with only linear optics. The protocol can be divided in two phases: initialization phase and preparation phase. We begin by describing the first one.

\subsection{Initialization phase}

We use spontaneous parametric down-conversion to prepare an entangled pair of photons (see Fig.~\ref{fig:init}). The initial state of these photons can be written as
\begin{equation}
|\psi_{}\rangle = \frac{1}{\sqrt{2}} \left ( |HV\rangle + |VH\rangle \right),
\label{triplet}
\end{equation}
where H and V denote state of single photon polarized horizontally or vertically, respectively. Using two polarizing beam splitters that reflect vertical and transmit horizontal polarization we are able to obtain the four-mode state
\begin{equation}
|\psi_{}\rangle = \frac{1}{\sqrt{2}} \left ( |1001\rangle + |0110\rangle \right ),
\label{stav_po_pbs}
\end{equation}
where the Fock state basis is used. Four numbers in each bracket denote the number of photons found in spatial modes $1$ through $4$, respectively. Half-wave plates HWP1 and HWP2 unify polarization states in all four modes. Then the Hong-Ou-Mandel interference \cite{bib:hom} 
takes place on balanced beam splitters BS1 and BS2 and the state of the photons 
changes to 
\begin{equation}
|\psi_{}\rangle = \frac{1}{2} \left ( |2000\rangle + |0200\rangle + |0020\rangle + |0002\rangle \right ).
\label{4x2000}
\end{equation}
If we want to generate a three-mode state, then we simply omit the last mode. Since in our protocol we post-select cases when two photons are present at the output of the preparation device, 
we can assume the effective initial two-photon state in the form 
\begin{equation}
|\psi_{}\rangle = \frac{1}{\sqrt{3}} \left ( |200\rangle + |020\rangle + |002\rangle \right ),
\label{eq:3x200}
\end{equation}
which will be used in the second phase of the protocol.

\subsection{Preparation phase}

In order to prepare an arbitrary state  (\ref{eq:target}) the state (\ref{eq:3x200}) is fed into a multiport linear interferometer schematically illustrated in Fig.~\ref{fig:scheme}. The device is composed of passive linear optical elements such as beam splitters and phase shifters and involves  filters F1 and F2 where the signal beams are attenuated by mixing them with auxiliary vacuum beams on beam splitters. The filtering operation is successful if the photons do not leak into the output auxiliary ports. This can be, in principle, verified by means of single-photon detectors placed on the auxiliary outputs, as shown in Fig.~\ref{fig:scheme}.  
However, such verification would require perfect detectors with unit efficiency. Instead, we can utilize a simpler verification strategy relying on postselection. To prove that the scheme is working, one can set detectors on every output mode and look for double coincidences which herald successful preparation of the state. Quantum state tomography \cite{bib:paris:estimation} could be used to completely characterize the generated state. In this case the target state is destroyed by the measurement. If the prepared state  serves as an input for another scheme, one can post-select successful preparations at the output of such scheme.

The parameters of the interferometer elements can be determined by ``reverse engineering'' 
of the target state. We let the target state propagate through the scheme in reverse and we successively get rid of all terms but the terms that are present in the state (\ref{eq:3x200}). The reversal is easily done for beam splitters (BS1, BS2 and BS3) and phase shifters (PS1, PS2 and PS3) because they are represented by unitary transformations. 
Filters F1 and F2 are also (probabilistically) reversible. Once the reverse procedure finds correct parameters for all elements, the scheme is designed to generate the target state (\ref{eq:target}) from the state (\ref{eq:3x200}).

In the Heisenberg picture, passive linear transformations can be described as linear unitary transformations of the creation operators, c.f. Eq. (\ref{eq:a_general_transform}). Every three-mode transformation can be decomposed into a sequence of two-mode and single-mode transformations  corresponding to beam splitters and phase shifters \cite{bib:reck:operator}. A fully general interferometric scheme can thus be constructed from these basic building blocks. Such a device
would provide a large number of degrees of freedom that can be exploited to optimize the success probability of the state preparation but at the same time its experimental realization  would be quite difficult. We have therefore chosen some sort of compromise. 
Our scheme includes all necessary elements needed for the preparation of arbitrary state (\ref{eq:target}) and one additional beam splitter BS3 and phase shifter PS4 as optional optimization elements. These components can be removed and our scheme thus simplified at the expense of a reduced probability of successful state preparation.

In what follows we imagine that the state propagates backwards through the scheme from the right to the left. We successively eliminate all terms where two photons are present in two different spatial modes and end up with the state (\ref{eq:3x200}) as a result.
First of all we get rid of the cross term $|101\rangle$ with complex amplitude $\epsilon$ by destructive interference. To this end, we employ a unitary transformation U101 represented by a beam splitter BS2 with amplitude transmittance $\tau_2$ accompanied by phase shifter PS3 imposing phase shift $\phi_3$. 
By a simple calculation, one finds that setting $\phi_{3} = - \arg \delta + \arg \epsilon$
and $\tau_2 = \frac{u}{\sqrt{1+u^2}}$, where $u = \left | \frac{\epsilon}{\delta} \right |$, the state $|101\rangle$ is completely eliminated by destructive interference.

The success rate of the protocol can be increased by the optimization beam splitter BS3 mixing modes $1$ and $2$. Amplitude transmissivity $\tau_3$ of BS3 should be chosen such that in the reverse propagation after transformation U101 the amplitude of the state $|110\rangle$  is minimized. The optimal value of $\tau_3$ can be found numerically by a simple computer algorithm that maximizes the overall probability of success of the protocol. The reverse propagated state impinging on the filter F2 from the right 
can be written as
\begin{equation}
|\psi^\prime\rangle =  \alpha' |200\rangle + \beta' |020\rangle + \gamma' |002\rangle +   +\delta'|110\rangle+\eta'|011\rangle.
\label{stateprime}
\end{equation}

If the term $|110\rangle$ is not completely eliminated and $\delta' \neq 0$ we are forced to employ a filter F2 in modes $1$ and $2$. Let $\hat{a}^\dag$ and $\hat{b}^\dag$ denote the creation operators for modes $1$ and $2$, respectively. The filtration by F2 can be expressed as
\begin{equation}
\left (
\begin{array}{c}
\hat{a}^\dag_{\mathrm{out}} \\ \hat{b}^\dag_{\mathrm{out}}\\
\end{array}
\right )
= 
\left (
\begin{array}{cc}
q_2 & -\frac{\delta' q_2}{\alpha' \sqrt{2}}  \\ 0 & q_2 \\
\end{array}
\right )
\left (
\begin{array}{c}
\hat{a}^\dag_{\mathrm{in}} \\ \hat{b}^\dag_{\mathrm{in}}\\
\end{array}
\right ),
\label{F1a}
\end{equation}
where $q_2$ is set so that the matrix can make part of a bigger unitary transformation on three modes. Without loss of any generality we can assume that $\delta'/\alpha'$ is real and positive because the phase shift between the two amplitudes is compensated by the phase shifter PS2 if we set 
 $\phi_{2} = \arg \alpha - \arg \epsilon$. The amplitude transmittance of the beam splitters forming the filter F2 is equal to $q_2$. The correct mixing of modes is achieved if the product of the amplitude reflectances $1-q_2^2$ of the two beam splitters is equal to $-\delta' q_2/\sqrt{2}\alpha'$.  This yields a quadratic equation for $q_2$ whose solution reads
\begin{equation}
q_2=\frac{1}{2}\left[\frac{\delta'}{\sqrt{2}\alpha'}-\sqrt{\left(\frac{\delta'}{\sqrt{2}\alpha'}\right)^2+4}\right].
\label{q2}
\end{equation}
The filtering requires an ancilla vacuum mode and the transformation (\ref{F1a}) is effectively implemented if no photon leaks into the output ancilla. Note that
Eq.~(\ref{F1a}) does not preserve canonical commutation relations because the operator of ancilla mode is omitted. This simplified mathematical description of the filter is nevertheless correct for our purposes. 

After filter F2 the state simplifies to 
\begin{eqnarray}
|\psi^{\prime\prime}\rangle =  \alpha'' |200\rangle + \beta'' |020\rangle + \gamma'' |002\rangle +  \eta'' |011\rangle.
\label{psitwoprime}
\end{eqnarray}
To ensure correct state preparation, one has to compensate filtering of modes $1$ and $2$ by additional attenuation of the third mode by a beam splitter with amplitude transmissivity $q_2$. The probability of success $P_{F2}$ of the filter $F2$ can be lower bounded as follows. The successful filtration during the state preparation procedure can be described by the transformation $|\psi'\rangle=M|\psi''\rangle,$ where the non-unitary operator $M$ depends on $q_2$.
We have $P_{F2}=\langle \psi''|M^\dagger M |\psi''\rangle/\langle \psi''|\psi''\rangle$ from which we obtain the inequality $P_{F2} \geq m_{\mathrm{min}}$ where $m_{\mathrm{min}}$ is the lowest eigenvalue of the matrix $M^\dagger M$. A straightforward calculation yields
\begin{eqnarray}
m_{\mathrm{min}}&= &\frac{1}{2}[q_2^4+(1-q_2^2+q_2^4)^2] \nonumber \\
 & & -\frac{1}{2}(1-q_2^2)^2\sqrt{1+6q_2^4+q_2^8}.
 \label{mmin}
\end{eqnarray}
It can be shown that $m_{\mathrm{min}}$ is a monotonically increasing function of $q_2$ which suggests that in order to maximize the success rate of the protocol we should maximize $q_2$.
It follows from Eq. (\ref{q2})  that $q_2^2$ monotonically decreases as the ratio $\delta'/\alpha'$ increases. It is therefore desirable to minimize $\delta'$and  maximize $\alpha'$ which is precisely the purpose of the beam splitter BS3. 
Let us now derive a lower bound on $q_2^2$. The maximum of absolute values of the amplitudes specifying the state (\ref{eq:target}) is greater or equal to $1/\sqrt{6}$. If the maximum amplitude corresponds to one of the states with two photons in a single mode then by simple relabeling of the modes we get $\alpha' \geq 1/\sqrt{6}$. If this amplitude corresponds to mode with two photons in two different modes, say $|110\rangle$ after possible re-labeling,
then we may use balanced beam splitter BS3 together with proper phase shifts to reach amplitude 
$\alpha' \geq 1/\sqrt{12}$. Since the unitary operations U(BS3) and U101 are deterministic, the state (\ref{stateprime}) is normalized and $|\alpha'|^2+|\delta'|^2 \leq 1$ which implies
$|\delta'|\leq \sqrt{\frac{11}{12}}$. Taking everything together, we find that 
$\left|\delta'/\alpha'\right| \leq \sqrt{11}$ can always be satisfied so that 
$q_2^2 \geq (15-\sqrt{209})/4\approx 0.136$ holds. If we plug in this lower bound into Eq. (\ref{mmin})
we finally get $P_{F2}\geq 0.52\%$.

In the next step of the protocol, the term $\eta'' |011\rangle$ in the state (\ref{psitwoprime}) is  removed by a destructive interference  between second and third modes. This is accomplished by a  unitary transformation U011 which consists of  a phase  shifter PS1 imposing phase shift $\phi_1$  on the second mode and a beam splitter BS1 with transmittance $\tau_1=\cos\vartheta$ mixing the second and third mode. 
The term $\eta''|011\rangle$ is eliminated provided that 
\begin{eqnarray}
\tan \phi &=&  \frac{\mathrm{Im}[(\gamma''-\beta'')/\eta'']}{\mathrm{Re}[(\gamma''+\beta'')/\eta'']},
\nonumber \\
\tan (2\vartheta) &=& \frac{\sqrt{2}\eta''}{e^{-i\phi}\gamma''-e^{i\phi}\beta''} , 
 \nonumber \\
\label{eq:condition_u011}
\end{eqnarray}
After the whole procedure we obtain the state 
\begin{eqnarray}
|\psi'''\rangle = \alpha''' |200\rangle + \beta''' |020\rangle + \gamma''' |002\rangle,
\label{psithree}
\end{eqnarray}
which can be transformed by filters F1  into the state (\ref{eq:3x200}). We denote by  $P_{F1}$ 
the overall success probability of filters F1. In order to obtain the state (\ref{psithree}) from the initial state (\ref{eq:3x200}) we in fact need to apply attenuation filters to only two modes. With probability $\frac{1}{3}$ the two photons are present in a mode which is not attenuated by F1 and, consequently, it  holds that $P_{F1} \geq \frac{1}{3}$.

\begin{figure}[!t!]
\begin{center}
\scalebox{1}{\includegraphics{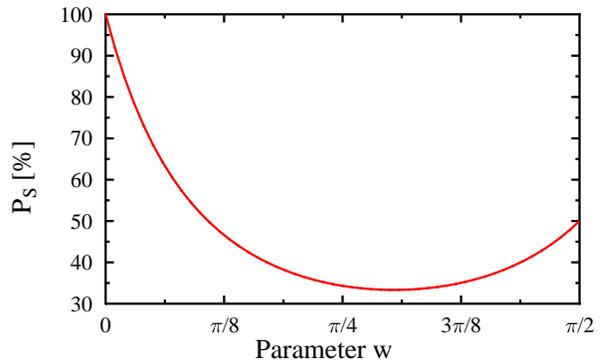}}
\end{center}
\caption{Success probability $P_S$ of the  protocol for states (\ref{test_1}) is plotted as a function of the parameter $w \in [0;\frac{\pi}{2}]$ which specifies the 
``difference'' of the target state from the initial state (\ref{eq:3x200}).}
\label{fig:test_amp}
\end{figure}

\begin{figure}[!t!]
\begin{center}
\scalebox{1}{\includegraphics{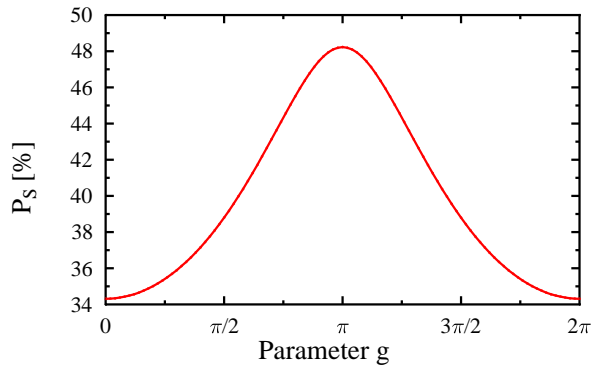}}
\end{center}
\caption{Success probability $P_S$ of the protocol for states (\ref{test_2}) is plotted as a function of the phase shift $g\in[0,2\pi].$}
\label{fig:test_phase}
\end{figure}

Parameters of all components are now determined. By proper inversion of the filters we obtain configuration which  prepares the target state (\ref{eq:target}) from the state (\ref{eq:3x200}). 
The total success probability for the whole protocol is defined as
\begin{equation}
P_S = \langle \psi_{\mbox{\tiny{OUT}}} | \psi_{\mbox{\tiny{OUT}}}\rangle, 
\label{final_prob}
\end{equation}
where $|\psi_{\mbox{\tiny{OUT}}} \rangle$ is un-normalized output state obtained from the normalized input state (\ref{eq:3x200}). It holds that $P_S =P_{F1} P_{F2} \geq 0.17\%$.

\section{Numerical simulations}

To verify the functionality of our protocol, we have performed  extensive numerical simulations of preparation of various states. An example of the results is given in Fig.~\ref{fig:test_amp} which displays the probability of successful preparation for target states
\begin{eqnarray}
|\psi_{1}\rangle & \propto &  \cos w \, (|200\rangle + |020\rangle + |002\rangle) 
\nonumber \\
 && + \sin w \, ( |110\rangle + |101\rangle + |011\rangle) ,
\label{test_1}
\end{eqnarray}
where $w \in[0,\frac{\pi}{2}]$. We can see that as $w$ increases the target state is ``shifting off'' the state (\ref{eq:3x200}) which is the product of the initialization phase. 
Even though the probability decreases, it is still well above $30\%$. It can be shown that for this particular class of states the best performance of the protocol is obtained by using a balanced optimizing beam splitter BS3. This probability $P_S$ is however very sensitive to exact splitting ratio of the beam splitter when $w \approx 0.3 \pi$. It this case it is therefore more convenient for the  experimental realization to use an unbalanced beam splitter with splitting ratio that corresponds to another local maximum of the success probability. Even though this choice gives a slightly lower success probability, it makes the scheme more robust and  tolerant to imprecision in the splitting ratio.

In another  numerical simulation we have studied preparation probability $P_S$ for a more asymmetric single-parametric class of states
\begin{eqnarray}
|\psi_{2}\rangle &\propto& |200\rangle + |101\rangle + |002\rangle 
\nonumber \\
& &+ e^{g\mathrm{i}}(|110\rangle + |020\rangle + |011\rangle),
\label{test_2}
\end{eqnarray}
where $g\in[0,2\pi]$. The result (see Fig.~\ref{fig:test_phase}) shows a symetric Gauss-like function realising maximum of $48\%$ at $g=\pi$ and asymptotically aproaching minimum of $34\%$ for $g=0,2\pi$.

\begin{figure}[!t!]
\begin{center}
\includegraphics{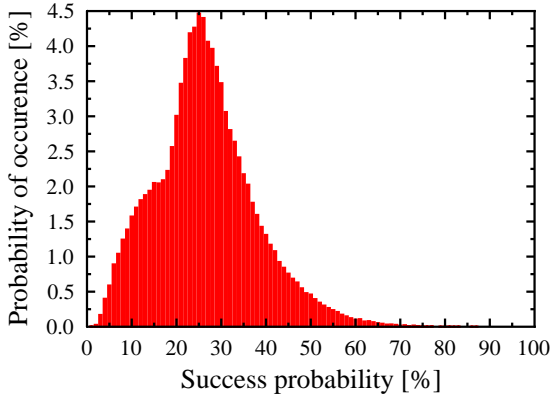}
\end{center}
\caption{Success probability histogram. For a given randomly generated set of $10^5$ states, the minimum success probability of the scheme reads $0.69\%$. The average success probability is around $26.0\%$.}
\label{fig:hist}
\end{figure}

\begin{figure*}[!t!]
\centerline{\psfig{figure=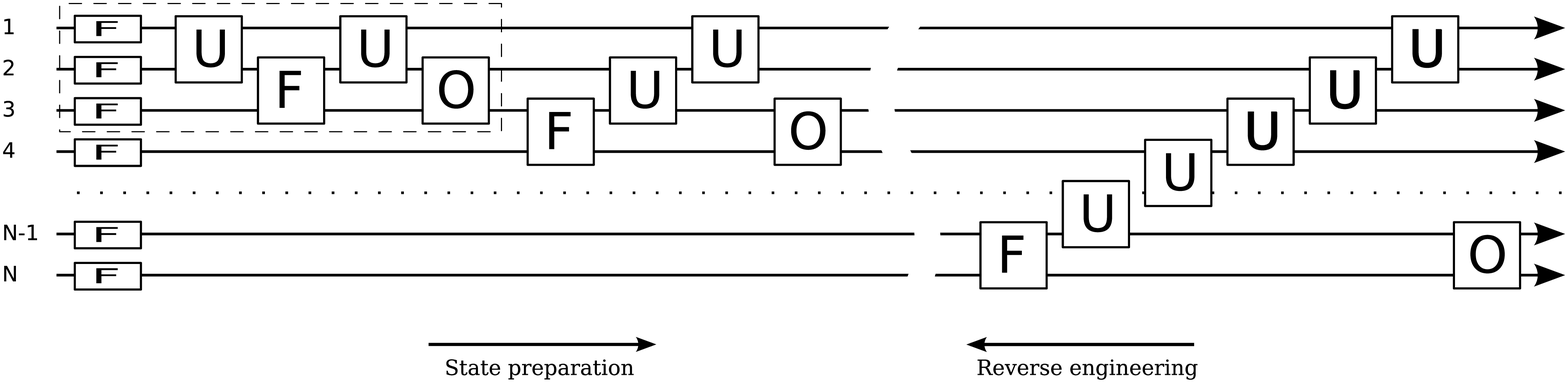,width=0.95\linewidth}}
\caption{Scheme for preparation of  generic $N$-mode two-photon state. As explained in the text, the scheme works in cycles. The scheme is composed of unitary transformations U that consist of a beam splitter and two phase shifters before and after the beam splitter, filters F  equivalent to filter F2 from the scheme in Fig.~\ref{fig:scheme}, and the optimizing beam splitters O.}
\label{fig:gener}
\end{figure*}

Finally, we present in  Fig.~\ref{fig:hist} a histogram of the success probability $P_S$ of the preparation protocol. This histogram was obtained by evaluating $P_S$ for $10^5$ randomly generated states. We can see that most of the states can be prepared with probability larger than $15\%$. However, there is a small class of states that are hard to generate. The lowest success probability obtained by the numerical calculations reads $0.69\%$ which qualitatively agrees with the lower bound $ 0.17\%$ derived in the preceding section.

\section{Generalization to an arbitrary number of modes}

So far we have explained how linear optics can be used to prepare an arbitrary state of two photons in three spatial modes. We now generalize our procedure to an arbitrary number of modes. Figure~\ref{fig:gener} schematically depicts the scheme for preparation of an arbitrary two-photon state of the form
\begin{equation}
|\psi_{\mbox{\footnotesize{target}}}\rangle = \sum_{j=1}^{N}\alpha_j|2_j\rangle + \sum_{j,k=1, j\neq k}^{N}\beta_{jk}|1_j1_k\rangle.
\label{Nmodestate}
\end{equation}
Here $N$ denotes the number of modes, $|2_j\rangle$ represents state with two photons in $j$th mode and vacuum in all other modes and $|1_j1_k\rangle$ indicates state with one photon in modes $j$ and $k$ an no photons in the other modes.  

We begin by employing non-degenerate spontaneous parametric down-conversion to produce
correlated state of two photons in $d=N/2$ pairs of modes $ja$ and $jb$,
\begin{equation}
|\psi_{\mathrm{SPDC}}\rangle=\frac{1}{\sqrt{d}} \sum _{j=1}^{d} |1_{ja}\rangle|1_{jb}\rangle.
\end{equation}
The modes $ja$ and $jb$ can be for instance different directions on the cone of the emission from the crystal \cite{bib:cinelli:all_versus_nothing}. Alternatively, one can pump the non-linear crystal with a sequence of $d$ ultra-short pulses and then the modes $ja$ and $jb$ correspond to $j$th time-bin \cite{bib:stucki:two-photon}. Yet another option is to simultaneously pump $d$ nonlinear crystals. After down-conversion, the pairs of modes $ja$ and $jb$ are combined on $d$ balanced beam splitters and the Hong-Ou-Mandel interference produces the state 
\begin{equation}
|\psi_{\mathrm{in}}\rangle=\frac{1}{\sqrt{N}} \sum _{j=1}^{N} |2_j\rangle.
\label{noon}
\end{equation}

In the second part of the protocol, the $N$-mode two-photon state (\ref{noon}) is 
injected into the interferometer shown in Fig.~\ref{fig:gener}. 
The parameters of the interferometer can be again determined by reverse engineering where we let the target state propagate backwards through the scheme and at each step we eliminate certain component of the state by destructive interference. In this way we finally obtain the state (\ref{noon}).
We now describe the cycles of the protocol in more detail. In the first cycle, we get rid of all states with exactly one photon in the last mode. First we apply unitary transformation to the first and second mode. This transformation is equivalent to the transformation U101 from Fig.~\ref{fig:scheme} and  eliminates the term $|100...001\rangle$. Next the second unitary transformation acting on the second and third mode eliminates the term $|010...001\rangle$. We continue in this fashion with only one exception. We do not use unitary transformation to mix the ($N$-1)th and $N$th mode. Instead, we need to use a filter F similar to the filter F2 shown in Fig.~\ref{fig:scheme}, accompanied by attenuation of the first $N-2$ modes (not shown in the figure). This filter removes the term $|00...11\rangle$. At this stage, all terms with exactly one photon in the last ($N$th) mode are eliminated and there is either zero or two photons in the last mode. For subsequent considerations it can be therefore neglected  and we have effectively reduced the $N$-mode problem to ($N$-1)-mode problem.

In the second cycle we apply the same strategy to eliminate all cross terms with one photon in the mode $N-1$. Repeating this procedure $N-3$ times we finally get back to the three-mode situation which we have already solved. The success probability of the protocol can be improved by employing optimizing beam splitters before each cycle. These beam splitters play the same role as  BS3 in Fig.~\ref{fig:scheme} and they should combine the modes where the filter will be applied in a given cycle. The number of required two-mode operations that have to be implemented scales as $\frac{1}{2}N(N-1)$ with $N$ being the number of modes.

\section{Preparation of two-photon KLM states}

The general scheme presented in the previous section can prepare any two-photon quantum state in an arbitrary number of modes. The experimental difficulty however increases considerably as the number of modes grows. If one is interested in preparation of specific class of quantum states then one could attempt to design a less general but simpler and experimentally more feasible scheme for this purpose. 
As an example of an important specific class of states we consider here  the so-called two-photon four-mode KLM states,
\begin{equation}
|\psi_{\mbox{\footnotesize{KLM}}}\rangle = \alpha |1100\rangle + \beta |0110\rangle + \alpha |0011\rangle.
\label{eq:klm}
\end{equation}
These states were introduced by Knill, Laflamme, and Milburn in the context of quantum computing with linear optics  \cite{bib:knill:klm}.  They can be used for teleportation-based implementation of quantum gates and the teleportation fidelity can be optimized by appropriate tuning  of $\alpha/\beta$ ratio \cite{bib:franson:klm}. In view of these applications it is clearly important to be able to prepare such states in the simplest possible way.

\begin{figure}[!t!]
\begin{center}
\includegraphics{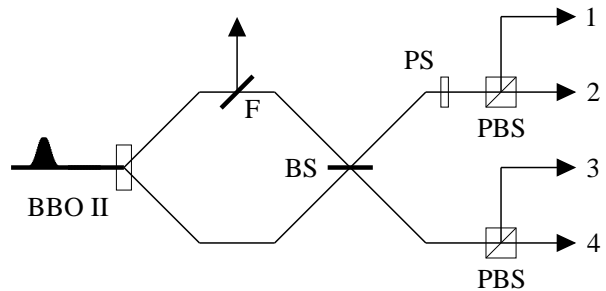}
\end{center}
\caption{Scheme for preparation of the two-photon four-mode KLM states.}
\label{fig:klmPrep}
\end{figure}

\begin{figure}[!t!]
\begin{center}
\includegraphics{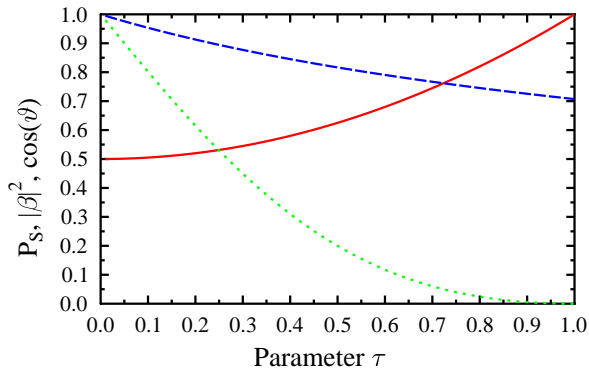}
\end{center}
\caption{The success probability of the protocol (full line), the probability of finding the state $|0110\rangle$ in the KLM state (dotted line) and the  amplitude transmissivity  of the beam splitter BS (dashed line) are plotted as functions of the parameter $\tau$.}
\label{fig:klmSimul}
\end{figure}

The scheme for preparation of the two-photon KLM states is shown in Fig.~\ref{fig:klmPrep}. 
A strong coherent laser beam pumps nonlinear crystal BBO cut for Type-II phase matching 
where a maximally entangled two-photon triplet state $|\psi\rangle = \frac{1}{\sqrt{2}} (|H_1V_2\rangle + |V_1H_2\rangle$ is generated in the process of spontaneous parametric down-conversion (indexes 1 and 2 denote the spatial modes). The first spatial mode is then subjected to polarization sensitive filtering using a beam splitter F which entirely transmits vertical polarization and has amplitude transmissivity $\tau$ for horizontal polarization. In practice, such filtering can be accomplished, e.g., by means of tilted glass plates \cite{bib:cernoch2006}. The two spatial modes are then combined on an ordinary (polarization in-sensitive) unbalanced beam splitter BS. We parametrize the transmittance $t$ and reflectance $r$ of BS by an angle  $\vartheta$ and we have $t=\cos\vartheta$ and $r=\sin\vartheta$.
In order to generate the desired state (\ref{eq:klm}) we have to set
\begin{equation}
\vartheta = \arctan{\sqrt{\tau}}.
\end{equation}
The polarization sensitive phase shifter PS adds a phase shift of $\pi$ to the vertically polarized photon in the first output spatial mode. As a result of this entire procedure  we obtain the state
\begin{equation}
|\psi\rangle = \alpha |H_1V_1\rangle + \beta |H_1V_2\rangle + \alpha |H_2V_2\rangle,
\end{equation}
which can easily be split by two polarizing beam splitters to obtain the state (\ref{eq:klm})
with parameters
\[
\alpha=\sqrt{\frac{\tau}{1+\tau^2}}, \qquad 
\beta=\frac{1-\tau}{\sqrt{1+\tau^2}}.
\]
Since the  scheme contains a filter F the state is prepared only probabilistically and the success probability is equal to $P_{S}=\frac{1}{2}(1+\tau^2)$.
Figure~\ref{fig:klmSimul} illustrates the trade-off between success probability of the whole protocol and the probability $|\beta|^2$ of finding $|0110\rangle$ in the generated state.
The teleportation protocol proposed by KLM  requires the state (\ref{eq:klm}) with $\alpha = \beta$  and for this case we find $P_S=0.57$.

\section{Conclusions}

In this paper we have demonstrated  how entangled pairs of photons generated in the process of spontaneous parametric down-conversion can be used for preparation of an arbitrary multimode two-photon state. This preparation requires only linear optical elements (beam splitters, wave plates, phase shifters) whose parameters can be determined analytically by reverse engineering procedure whereby the target state is propagated backwards through the scheme.  We have also suggested a simple and experimentally feasible setup specifically tailored for preparation of  two-photon KLM states which are crucial for linear optics quantum computing. The proposed schemes rely only on existing and well mastered experimental techniques and can be therefore implemented with current technology.
Our findings thus pave the way towards generation of complex multimode entangled states of light required for advanced quantum information processing, ultraprecise measurements or quantum lithography.

\acknowledgments
This research was supported by the Ministry of Education of the Czech
Republic under the project Centre of Modern Optics (LC06007).

\end{document}